\documentclass[%
 aip,
 amsmath,amssymb,
 reprint,%
]{revtex4-1}

\usepackage{graphicx}
\usepackage{dcolumn}
\usepackage{bm}

\usepackage[utf8]{inputenc}
\usepackage[T1]{fontenc}
\usepackage{mathptmx}
\usepackage{etoolbox}

\usepackage[colorinlistoftodos]{todonotes}

\usepackage[colorinlistoftodos]{todonotes}

\usepackage[colorinlistoftodos]{todonotes}

\makeatletter
\def\@email#1#2{%
 \endgroup
 \patchcmd{\titleblock@produce}
  {\frontmatter@RRAPformat}
  {\frontmatter@RRAPformat{\produce@RRAP{*#1\href{mailto:#2}{#2}}}\frontmatter@RRAPformat}
  {}{}
}%
\makeatother
\begin{document}

\preprint{AIP/123-QED}

\title[Exciting high-frequency short-wavelength spin waves]{Exciting high-frequency short-wavelength spin waves using high harmonics of a magnonic cavity mode}

\author{Nikhil Kumar}
\email{nikhilkumarcs@nitc.ac.in}

 \affiliation{Electronics and Communication Engineering Department, National Institute of Technology Calicut, 673 601 India}
\author{Paweł Gruszecki}%
 \email{gruszecki@amu.edu.pl}
\affiliation{ 
Institute of Spintronics and Quantum Information, Faculty of Physics, Adam Mickiewicz University, Poznań, Uniwersytetu Poznańskiego 2, 61-614 Poznań, Poland
}%

\author{Mateusz Gołębiewski}%
\email{mateusz.golebiewski@amu.edu.pl}
\affiliation{ 
Institute of Spintronics and Quantum Information, Faculty of Physics, Adam Mickiewicz University, Poznań, Uniwersytetu Poznańskiego 2, 61-614 Poznań, Poland
}%

\author{Jaros{\l}aw W. K{\l}os}
\email{klos@amu.edu.pl}
\affiliation{ 
Institute of Spintronics and Quantum Information, Faculty of Physics, Adam Mickiewicz University, Poznań, Uniwersytetu Poznańskiego 2, 61-614 Poznań, Poland
}%

\author{Maciej Krawczyk}
\email{krawczyk@amu.edu.pl}
\affiliation{ 
Institute of Spintronics and Quantum Information, Faculty of Physics, Adam Mickiewicz University, Poznań, Uniwersytetu Poznańskiego 2, 61-614 Poznań, Poland
}%

\date{\today}

\begin{abstract}
Confined spin-wave modes are a promising object for studying nonlinear effects and future quantum technologies. Here, using micromagnetic simulations, we use a microwave magnetic field from a coplanar waveguide (CPW) to pump a standing spin-wave confined in the cavity of magnonic crystal. We find that the frequency of the fundamental cavity mode is equal to the ferromagnetic resonance frequency of the plane film and overlaps with the magnonic bandgap, allowing high magnetic field tunability. Multi-frequency harmonics of the cavity mode are generated once the microwave amplitude surpasses a certain threshold. Specifically, the second and third harmonics at 0.5 T equate to 48.6 and 72.9 GHz with wavelengths of 44 and 22 nm respectively, which propagate into the crystal. This effect reaches saturation when the CPW covers the entire cavity, making the system feasible for realization. These processes show potential for the advancement of magnonics at high-frequencies and very short-wavelengths.
\end{abstract}

\maketitle

Magnonics, which uses spin waves (SWs) to transmit and process information, is one of the fastest developing areas of research in modern magnetism\cite{kruglyak2010magnonics, Mahmoud2020,Nikitov_2015,Chumak_2017,Barman_2021,Flebus2023}. It covers a wide frequency range from sub-GHz to tens of THz, with corresponding wavelengths from micrometers to nanometers, and is free of translational electron motion and Joule heating. This makes it a promising alternative to electronics and photonics in terms of high operating frequency, low power consumption and miniaturization\cite{Csaba2017,Mahmoud2020,dieny2020opportunities,Hirohata2020,Pirro2021,chumak2022}. Magnonic crystals (MCs) are the magnetic equivalent of photonic or phononic crystals, which are important elements of photonic \cite{Joannopoulos08,Soljacic2004} and phononic \cite{Maldovan2013, Vasileiadis2021} technologies, respectively, and are also considered to be an important element of future magnonic circuits\cite{krawczyk2008plane,Lenk2011,Krawczyk_2014,chumak2017magnonic}. The structures formed from the magnetic thin film by periodic patterning are so-called planar MCs. The characteristic in-plane sizes of the considered structures are in the range of tens or single hundreds of nanometers, while the thicknesses of the magnetic films are about a few or tens of nanometers \cite{sokolovskyy2011magnetostatic, mamica2019nonuniform, klos2012impact, klos2014influence, gubbiotti2010brillouin}. 
In this context, the SW modes in antidot lattices (ADLs), i.e, thin ferromagnetic films periodically structured with holes, have attracted considerable interest \cite{Vysotskii2005magnetostatic, yu2007size, pechan2005direct, mcphail2005coupling, kostylev2008propagating, ulrichs2010magnonic, hu2011field, tacchi2015universal} and are useful for controlling SW velocity\cite{neusser2011magnonic}, exploiting magnonic band gaps and diffraction\cite{Gieniusz2017,Golebiewski2020,Riedel2023}.

However, the fundamental obstacles in the design of magnonic devices are, among others, difficulties related to the excitation of SWs, especially of high frequency and short wavelengths.
To excite short SWs, i.e. with wavelengths below 100 nm, the use of microwave stripe lines \cite{Ciubotaru2016}, magnetic solitons (domain wall \cite{Woo2017} or vortex \cite{Dieterle2019}), grating couplers \cite{Yu2016} or inhomogeneity of the internal magnetic field \cite{Mushenok2017} has been proposed. The first approach requires ultra-narrow metallic strips that limit the transmitted power, the latter two are used at low frequencies that are limited by the natural oscillations of the magnetic solitons or ferromagnetic nanoelements. In the latter case, the frequencies are usually below or around the ferromagnetic resonance frequency of the respective bulk material.

By introducing a defect into the ADL, e.g. in the form of a filled hole (or a few filled holes), we can trap SW energy in this region whenever a resonant frequency of cavity falls into the magnonic band gap of the MC \cite{Kruglyak2006,Klos2013,Di2014,Morozova2015}. Such a magnonic cavity allows building-up of SW amplitude \cite{kumar2018resonant, kumar2022enhanced}, similar to electromagnetic waves in a photonic crystal cavity \cite{Joannopoulos08}, when externally pumped at the cavity resonant frequency. Finally, at sufficiently high wave amplitudes, nonlinear processes are initiated. This effect is widely studied and exploited in photonics cavities, in particular, to enhance the generation of 2$^\text{nd}$ and 3$^\text{rd}$ harmonics
\cite{Soljacic2004,Lalanne2008}. 

Recently, multi-frequency generation in magnonics has also been demonstrated \cite{Schultheiss2019,Koerner2022,Hula2022,Gross2022}. However, the frequency of the generated SW modes hardly exceeds 10~GHz. In particular, the results of time-resolved scanning transmission X-ray microscopy measurements demonstrate the generation of SW high harmonics, up to $7^\text{th}$ order, in ADL based on 50~nm thick Py (antidot diameter 450~nm and lattice constant 900~nm) \cite{Gross2022}. The observed conversion efficiency is between 12.4\% and 33.3\%, while the pumping microwave field frequency is only 0.93~GHz, so the $7^\text{th}$ order harmonic frequency is $6.51$~GHz. However, the modes involved in the process are not clearly identified.
 
This paper presents micromagnetic simulations of an ADL in a Py thin film with a L$_{3}$ cavity (3 antidots removed), pumped by a standard coplanar waveguide (CPW) (Fig.~\ref{fig:Fig1}). The SW amplitude of the fundamental cavity mode can easily exceed the nonlinearity threshold, resulting in high frequency harmonics. The fundamental cavity mode frequency corresponds to the ferromagnetic resonance (FMR) frequency of the uniform Py layer and this frequency is located within the magnonic bandgap of the ADL. This allows wide frequency tunability by the magnetic field: changing it from 200~mT to 1~T results in the 2$^\text{nd}$ and 3$^\text{rd}$ harmonics in the frequency ranges 28-80~GHz and 42-120~GHz, while keeping the SW wavelength below 62 and 40~nm, respectively. The presented mechanism of high-frequency SW generation demonstrates its general character, the feasibility of its experimental realization, and its usefulness for the generation of high-frequency short-wavelength SWs for a wide range of magnonic applications.

\section{\label{sec:level2}Device model}

\begin{figure}
\includegraphics[width=8.6cm]{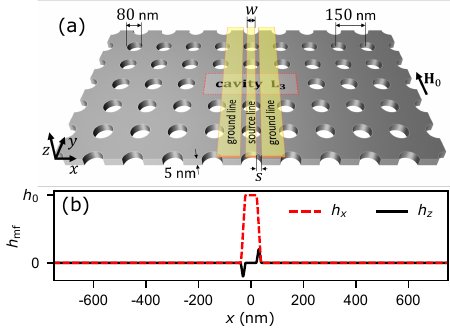}
\caption{
(a) Schematic of the device -- an antidot lattice with a three-hole defect cavity (L$_{3}$) centered at $x=0$ and $y=0$. Magnonic crystal is made of a 5~nm thick Py-film with magnetization saturated by the magnetic field $H_0$ directed along the $y$-axis, the diameter of the antidots is 80~nm and the lattice constant is 150~nm. A CPW antenna is placed just above the film, centered at $x_0$ (in the plot $x_0=0$), and consists of a source line of width $w$ and two ground lines, which are separated from the source line by gaps of width $s$. (b) The SWs are excited by the $x$ and $z$ components of the microwave frequency magnetic field ${\bf h}_{\text{mf}}$ of amplitude $h_\text{mf}$ produced by CPW. The calculated field distribution from CPW with $w=40$~nm and $s=20$~nm is shown.
}\label{fig:Fig1} 
\end{figure}

The device geometry used in this study is shown in Fig.~\ref{fig:Fig1}(a). It is an ADL with a square lattice of antidots in a 5~nm thick Py film, with an antidot diameter of 80~nm and a lattice constant of $a=150$~nm. A three-antidot defect L$_{3}$ (i.e., the area of 3 unit cells without the holes) is introduced in this structure. The system is uniformly magnetized by an in-plane bias magnetic field $\mu_0 H_0 >0.2$~T directed along the $y$-axis ($\mu_0$ is the permeability of the vacuum). The SWs are excited by a microwave frequency magnetic field ($\textbf{h}_\text{mf}$) generated by a CPW transducer consisting of a signal line of width $w$ separated by $s=20$~nm gaps from two identical ground lines (much wider than $w$), and with centered at $x_0$. To enhance the excitation of SWs, the CPW is aligned along the $y$-axis.

Two simulation methods are used to study the SW dynamics. Comsol Multiphysics package is used to calculate the magnonic band structure and cavity modes. Here, we use the finite element method (FEM) to solve the eigenproblem obtained from the linearized Landau-Lifshitz equation in real space and the frequency domain. The nonlinear effects are simulated in the real space and time domain (with finite difference time domain method, FDTD) using the Mumax3 micromagnetic solver\cite{vansteenkiste2014design}, which solves the full Landau-Lifshitz-Gilbert equation. The methods are described in detail in the Methods section. We use typical material parameters and typical values for Py, i.e., the saturation magnetization $M_{s}=800$~kA/m, the exchange constant $A=13$~pJ/m, and in FDTD simulations the damping $\alpha=0.01$. The magnetization dynamics inside the cavity are studied by probing the magnetization over $450\times150$~nm area, i.e., three unit cells along the $x$-axis and 1 along the $y$-axis.

\begin{figure}
\includegraphics[width=8.6cm]{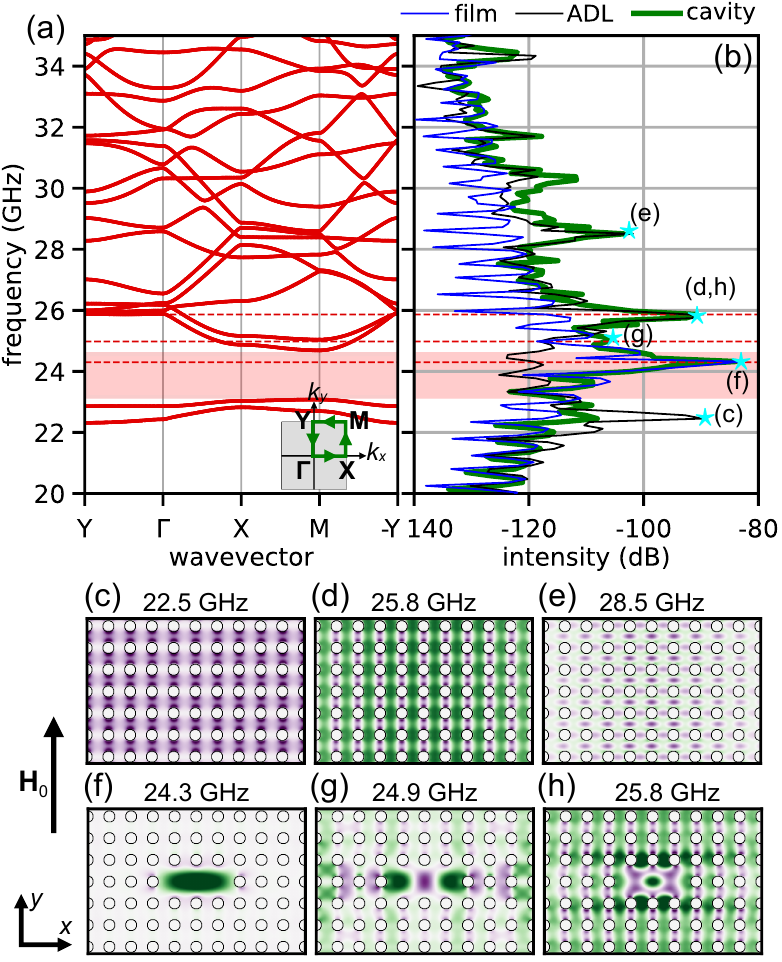}
\caption{
(a) Dispersion relation of SWs over the Y-$\Gamma$-X-M-Y path in the first Brillouin zone (as indicated in the inset) in the 2D ADL calculated with FEM. The magnonic band gap is marked in transparent red. (b) Spin-wave spectra obtained using FDTD using broadband excitation ($\mu_0 h_\text{mf}=1$~mT) by a spatially uniform microwave field for a uniform Py film (the blue line), an ADL (the black line), and an ADL with the L$_{3}$ cavity (the thick green line). The cyan stars indicate the mode profiles displayed in (c)-(h). Eigenmodes in the ADL are demonstrated in (c)-(e), while mode profiles for the ADL with L$_{3}$ cavity are presented in (f)-(h). The color utilized represents the amplitude of the out-of-the-plane component of the magnetization in arbitrary units. The external magnetic field $\mu_0H_0=0.5$~T is applied along the $y$-axis.
}\label{fig:Fig2} 
\end{figure}

\section{Results}
\subsection{Band structure and the resonance spectra}

The first step in our investigation is examining the magnonic band structure of the ADL (without the defect). It is shown in Fig.~\ref{fig:Fig2}(a) and is obtained by FEM at 0.5~T external magnetic field. The dispersion relation is over the path Y-$\Gamma$-X-M-Y in the first Brillouin zone (see the inset showing the first Brillouin zone and the high-symmetry path). Importantly, we see a well-defined full bandgap, ranging from 23.05~GHz to 24.95~GHz. The two bands below the band gap originate in the edge modes, i.e., the modes with the amplitude concentrated near the antidot edges, while the bands above are bulk modes, with the first one a fundamental mode of the ADL \cite{tacchi2015universal}.

The resonance response of the ADL, ADL with the L$_3$ cavity, and as a reference of the plain Py film are shown in Fig.~\ref{fig:Fig2}(b). The results are obtained with FDTD method by applying a spatially uniform broadband microwave magnetic field linearly polarized along the $z$-axis:
\begin{equation}
\mathbf{h}_{\text{mf}}=h_{\text{mf}}\text{sinc}\left(2\pi f_{\text{cut}}\left(t-t_{0}\right)\right)[0,0,1],\label{eq:spectra}
\end{equation}
where, the amplitude of microwave field: $ \mu_{0}h_\text{mf} = 1$~mT, the cut-off frequency: $f_\mathrm{cut}=40$~GHz, and $t_0=8/f_\mathrm{cut}$. 
The intensity is calculated by integrating the normalized $z$ component of the magnetization vector ($m_z\equiv M_z/M_\text{S}$) over the cavity region (i.e., the area of three unit cells in ADL and in the plane film) in a steady state. For presenting intensity, we use a logarithmic scale: $20\log_{10}(|m_z|)$ in dB units. 

For a plain Py film, we have only one distinctive peak at 24.3~GHz, corresponding to its FMR frequency. In the ADL there are several intense peaks \cite{tacchi2015universal}, the most intense are shown in Fig.~\ref{fig:Fig2}(c-d): (c) the symmetric edge mode at 22.5~GHz ($1^\text{st}$ band), (d) the fundamental excitation of the ADL at 25.8~GHz, and (e) a higher order SW at 28.5~GHz. This is in full agreement with the dispersion relation presented in Fig.~\ref{fig:Fig2}(a). For the ADL with the cavity, we obtained the spectrum (green line in Fig.2(b)) which is quite similar to the undetected ADL, but with a well-defined peak at $f_0=24.3$~GHz, i.e. in the bandgap of the ADL. The peak represents the fundamental cavity mode (see Fig.~\ref{fig:Fig2}(f)) which oscillates in phase, and its amplitude is confined inside the cavity. Interestingly, the frequency of this mode is the same as the FMR of a uniform Py film (see green and blue lines in Fig.~\ref{fig:Fig2}(b)). There are also high-frequency resonant modes in the cavity with enhanced SW amplitude inside the cavity, as shown in Fig.~\ref{fig:Fig2}(g) and (h). However, their frequencies overlap with the continuous bands of the ADL, and for that reason, the SW amplitude of these modes spread over the entire ADL.  

Learned from photonics, the enhancement of the pumping power on the cavity mode shall result in entering into nonlinear regime whenever the cavity is a medium possessing some kind of nonlinearity \cite{Soljacic2004,Lalanne2008}. In magnetism, nonlinearity is its inherent property~\cite{Gurevich1996}. Therefore, in the following part of the manuscript, we will focus on the fundamental cavity mode pumped by the microwave magnetic field generated by CPW at 24.3~GHz and at high amplitudes.

\subsection{Nonlinear dynamics}

\begin{figure}
\includegraphics[width=8.6cm]{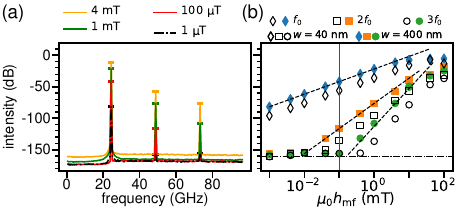}
\caption{
(a) The SW spectra ($\propto m_z^2$) in the L$_3$ cavity excited with CPW of $w=400$~nm, located at $x_0=0$ at frequency 24.3~GHz and amplitude $\mu_0 h_\text{mf}$: 1~\textmu T (the black-dashed line), 0.1~mT (the red line), 1~mT (the green line), and 4~mT (the orange line). (b) Intensity ($\propto m_z^2$) of the fundamental cavity mode ($f_0=24.3$~GHz), and its second ($f_2 \equiv 2f_0=48.6$~GHz), and third ($f_3 \equiv 3f_0= 72.9$~GHz) harmonics on the microwave excitation field strength $h_\text{mf}$ from CPWs with signal line widths $w=40$~nm and 400~nm represented as open and closed points, respectively. The black dashed lines represent the fitted dependencies from Eq.~(\ref{eq:intensity_dep}), whereas the vertical narrow green line represents the field of value 0.1~mT used in Fig.~\ref{fig:Fig4_v2}. Simulation with FDTD was performed at $\mu_0 H_0=0.5$~T.
}\label{fig:Fig3} 
\end{figure}

Frequency spectra of the ADL cavity system are shown in Fig.~\ref{fig:Fig3}(a). The calculations were performed for 0.5~T magnetic field and the CPW of $w=400$ nm wide, located at the center of the cavity ($x_0=0$), which generate the fields $h_{\text{mf}}=1$~\textmu T, 100~\textmu T, 1~mT, and 4~mT. Three peaks, corresponding to the frequencies $f_0=24.3$ GHz, $f_2=48.6$ GHz, and $f_3=72.9$ GHz, are clearly visible at higher amplitudes of the microwave field, except the lowest considered value $h_\text{ms} = 1$~\textmu T, where $f_2$ and $f_3$ are at the base signal level below $-150$~dB. These frequencies correspond to the excitation of the fundamental cavity mode ($f_0$) and its second ($f_2 \equiv 2f_0$) and third ($f_3 \equiv 3f_0$) harmonics, respectively. The higher the amplitude of the microwave field, the higher the intensities of the modes. However, there are significant changes in the relative peak amplitudes between the fundamental cavity mode and its harmonics. For the fundamental cavity mode, there is an increase of 70~dB in the peak as we increase the microwave amplitude from 1~\textmu T to 4~mT, whereas for the second and third harmonics, there is an increase of approximately 100~dB and 80~dB, respectively.

To shed more light on the efficiency of the SW excitations we perform systematic simulations with increased $h_{\text{mf}}$ from 1~\textmu T to 0.1~T using the same CPW antenna. The results are shown in Fig.~\ref{fig:Fig3}(b) with full diamond, square, and circle dots for $f_0$, $f_2$, and $f_3$, respectively. The intensity of the fundamental cavity mode increases from $-80$~dB at $10^{-3}$~mT till $-10$~dB at 10~mT, and saturates with a further increase of $h_{\text{mf}}$. The intensities of the 2$^\text{nd}$ and 3$^\text{rd}$ harmonics start to grow at about 0.01~mT and 0.1~mT, respectively, grow almost linearly in this double logarithmic scale, and both reach an intensity only 20~dB less than the $f_0$ mode at 40~mT. 

This means that the effectiveness of the nonlinear pumping of the second and third harmonic increases according to the square and cubic function of $h_{\text{mf}}$, with different threshold amplitudes of $h_{\text{mf}}$. This is according to the theoretical predictions for the plane film and plane waves \cite{Gurevich1996} indicating that the intensity of the $n^\text{th}$ harmonics:
\begin{equation} \label{eq:intensity_dep}
    I^{(n)} \propto \left(\frac{h_\text{mf}}{H_0}\right)^n.
\end{equation}
Similar behaviour has already been observed experimentally in the microwave pumping of the elliptical Py nanodot (10~nm thick and 1~\textmu m $\times$ 0.5~\textmu m lateral dimensions) saturated along the long axis \cite{Demidov2011}. However, in Ref.~[\onlinecite{Demidov2011}] the excitation was off-resonant, i.e., the generation of the 2$^\text{nd}$ and 3$^\text{rd}$ harmonic was only observed at $2f_\text{ex,1}$ and $3f_\text{ex,2}$, with the pumping field frequency $f_\text{ex,1}$ and $f_\text{ex,2}$, respectively. Here, $2f_\text{ex,1}$ and $3f_\text{ex,2}$ correspond to the frequency of the antisymmetric (with the nodal line along the long axis of the ellipse) and symmetric (fundamental) eigenmode of the nanodot, respectively. This means that for a given frequency of the microwave field $f_\text{ex,1}$ or $f_\text{ex,2}$ only the 2$^\text{nd}$ or 3$^\text{rd}$ harmonic can be observed. 

\begin{figure}
\includegraphics[width=8.6cm]{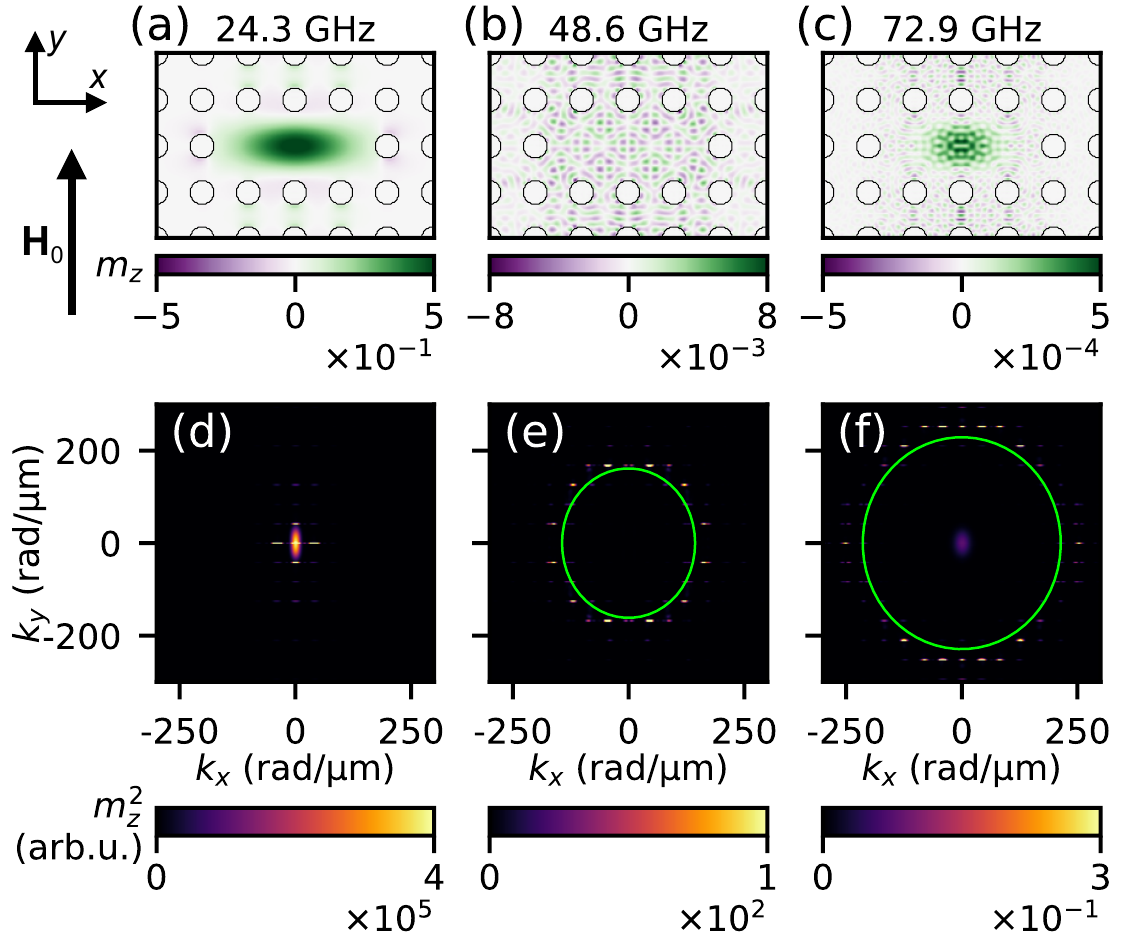}
\caption{
(a-c) Spatial distributions of amplitudes of the perpendicular component of reduced magnetization $m_z$ of the SW modes corresponding to the (a) fundamental cavity mode (24.3~GHz), and its (b) second (48.6~GHz), and (c) third (72.9~GHz) harmonics. FDTD results are obtained with the CPW of $w=400$~nm, $x_0=0$, $h_\text{mf}=4$~mT, and $\mu_0 H_0 = 0.5$~T. (d-f) Representation of the modes from (a)-(c) in the reciprocal space. The size of the Brillouin zone is ca. 42~rad/\textmu m. The green lines in (e) and (f) represent analytical\cite{kalinikos1986} isofrequency contours for uniform film at frequencies 48.6~GHz 72.9~GHz, respectively.
}\label{fig:Fig4} 
\end{figure}

The spatial distributions of the fundamental cavity mode ($f_0$), mode $f_2$ and $f_3$ at $h_\text{mf}=4.0$~mT microwave excitation field are shown in Fig.~\ref{fig:Fig4}(a), (b) and (c), respectively. The clear fine wavy pattern is seen at 48.6 and 72.9~GHz indicating a short wavelength of excited SWs. The amplitude of these harmonics spreads to the ADL, because, the accessible bands of the SWs in the ADL at these high frequencies are very dense with all the modes being propagating.
There are slightly larger amplitudes along the $y$-axis, which may be advantageous to use these modes to exploit the propagation of waves at these high frequencies. Indeed, Fig.~\ref{fig:Fig4}(e)-(f) shows that in reciprocal space the 2$^\text{nd}$ and 3$^\text{rd}$ harmonics have discrete spectra at high values of wavenumber. There is a fine regularly spaced arrangement of the spots in Fig.~\ref{fig:Fig4}(e)-(f) in the $k$-space, which is related to the periodicity of the structure and the discrete values of the reciprocal lattice vectors. These bright spots are located around the almost circular isofrequency contours of SWs in the plain Py film at 48.6 and 72.9~GHz. Their radius is 180 and 220~rad/\textmu m for $f_2$ and $f_3$, and correspond to the SWs at wavelengths as short as 35 and 29~nm, respectively. The spots along the direction of $H_0$ have larger intensities that indicate the preferred direction of SW propagation, accordingly with observation in Fig.~\ref{fig:Fig4}(b) and (c). In the future, the design of isofrequency contours of the ADL at relevant frequencies may allow the propagation direction of nonlinearly excited high-frequency SWs to be tailored to specific requirements. This can be achieved by proper design of the ADL geometry and the magnonic band structure~\cite{Klos2015,Sadovnikov2016}.

The spatial distributions of the fundamental cavity mode ($f_0$), mode $f_2$ and $f_3$ at a microwave excitation field of $h_\text{mf}=4.0$~mT are displayed in Fig.~\ref{fig:Fig4}(a), (b) and (c). The figures present a neat, undulating pattern at 48.6 and 72.9~GHz indicating a short wavelength of excited SWs. Notably, the amplitude of these harmonics extends from the cavity to the ADL as the magnonic bands in the ADL at these high frequencies are very dense with all the modes being propagating. Slight amplitude increases occur along the $y$-axis, suggesting that utilizing these modes could be useful for wave propagation at high frequencies. In fact, as seen in Fig.~\ref{fig:Fig4}(e)-(f), the 2$^\text{nd}$ and 3$^\text{rd}$ harmonics possess discrete spectra in reciprocal space with regularly spaced spots at high wavenumbers. This pattern is related to the structure's periodicity and the discrete values of its reciprocal lattice vectors. The bright spots are located around the almost circular isofrequency contours of SWs in the plain Py film at 48.6 and 72.9~GHz. For $f_2$ and $f_3$, their radius is 180 and 220~rad/\textmu m, respectively, which corresponds to the SWs at wavelengths as short as 35 and 29~nm. The spots have larger intensities along the direction of $H_0$ that indicate the preferred direction of SW propagation in the ADL, accordingly with observation in Fig.~\ref{fig:Fig4}(b) and (c).
In the future, designing isofrequency contours of the ADL at pertinent frequencies could enable tailoring of the propagation direction of nonlinearly excited high-frequency SWs towards specific needs. This objective can be accomplished through appropriate ADL geometry design and tailored magnonic band structure~\cite{Klos2015,Sadovnikov2016}.

\begin{figure}
\includegraphics[width=\linewidth]{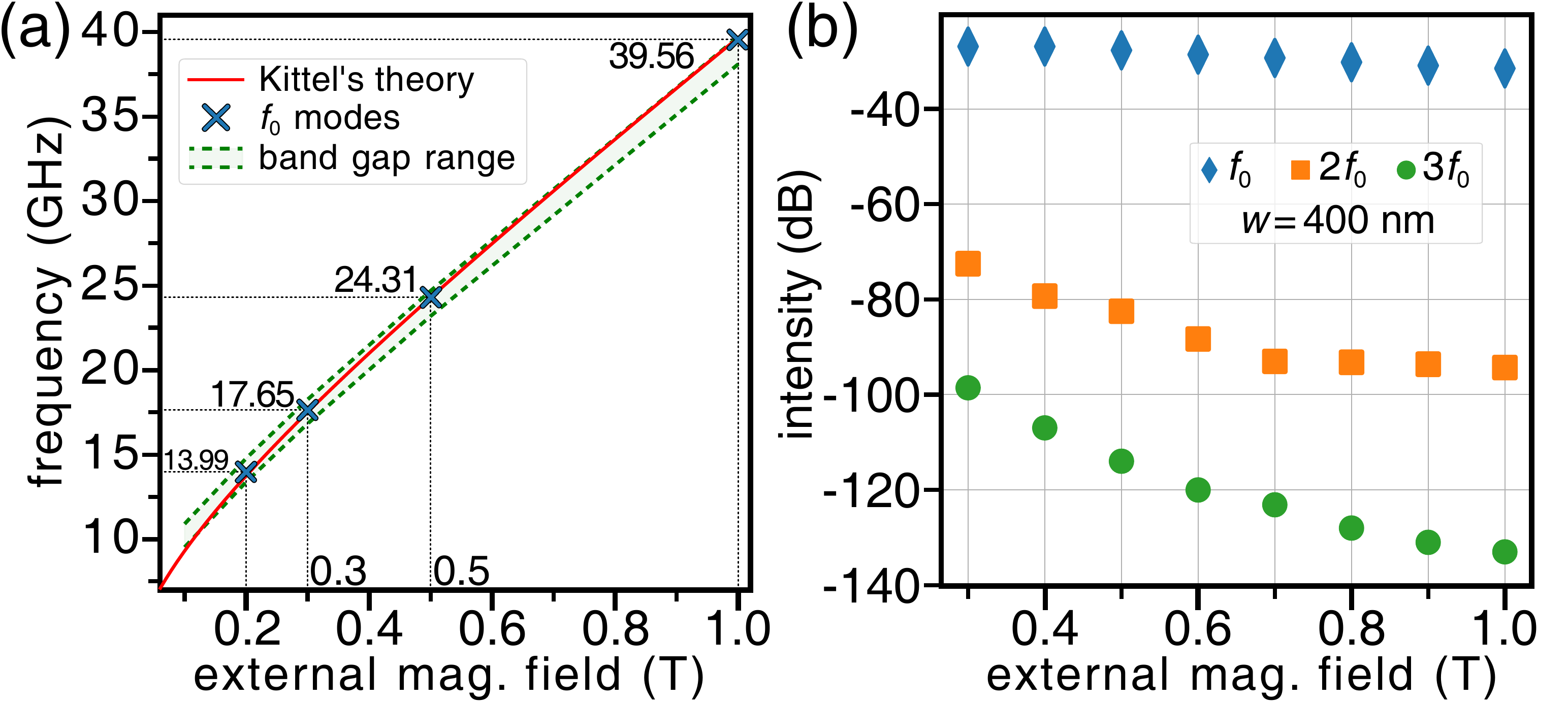}
\caption{(a) Dependence of the fundamental cavity mode frequency on external static magnetic field magnitude, $H_0$. The crosses mark the frequency of the cavity mode obtained from FEM in the supercell approach, while the red line represents the FMR frequency of the plain Py film from the Kittel's formulae \cite{kittel1946theory}. The green dotted lines indicate the edges of the band gap obtained from FEM.
(b) Intensity of the fundamental cavity mode ($f_0$) and its second ($2f_0$) and third harmonics ($3f_0$) as a function of $H_0$ obtained by FDTD. The markers indicate the intensities obtained by excitation with a CPW line $w=400$~nm wide at $\mu_0 h_\text{mf}=0.1$~mT at frequencies following the red line in (a).
}\label{fig:Fig6} 
\end{figure}

According to Eq.~(\ref{eq:intensity_dep}), the effectiveness of the nonlinear process can also be increased by decreasing $H_0$. However, to maintain similar conditions for the ADL cavity system at different values of $H_0$, the frequency of the fundamental cavity mode must follow the magnonic band gap as the magnetic field is varied. In Fig.~\ref{fig:Fig6}(a), we show that the frequency of the fundamental cavity mode follows the FMR frequency of the plain Py film, and in a wide range of fields, i.e. from 0.2 to 1~T, the frequency of this mode (increasing from 13.99 to 39.56~GHz, respectively) falls into the magnonic band gap. Taking the frequencies of the FMR mode with increasing $H_0$, assuming a pumped microwave magnetic field amplitude $h_\text{mf} = 0.1$~mT, we perform simulations. The results are shown in Fig.~\ref{fig:Fig6}(b). We see that as the magnetic field decreases from 1~T to 0.3~T, the intensity of the 2$^\text{nd}$ and 3$^\text{rd}$ harmonics increases by 21.7~dB and 34.5~dB respectively. However, there is a downside to decreasing $H_0$, it is that a lower frequency means longer wavelengths. In particular, at 1~T the 2$^\text{nd}$ and 3$^\text{rd}$ harmonics are at 79.12 and 118.68~GHz with wavelengths of 31.1 and 21.6~nm respectively, while at 0.2~T they are at 27.98 and 41.97~GHz and wavelengths of 61.4 and 39.8~nm, respectively.

\subsection{Getting ready to experiment}

The type of the ADL considered in this paper has been broadly investigated in the last few years \cite{neusser2008spin,neusser2011magnonic,tacchi2015universal,Zelent2017,Gross2022}, so the nonlinear effects presented above should be feasible for experimental realization. To make our idea closer to realization we studied the influence of the CPW width and position with respect to the cavity center on the frequency multiplication.

\begin{figure}
\includegraphics[width=8.6cm]{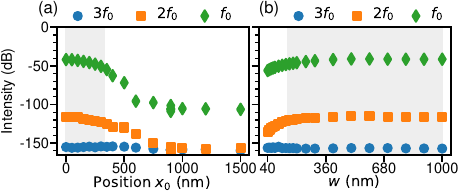}
\caption{
The dependence of the SW intensities ($\propto m_z^2$) at $f_0=24.3$~GHz, $f_2=2\times 24.3$~GHz, and third $f_3=3\times 24.3$~GHz on (a) the position of the CPW antenna $x_0$ with respect of cavity center ($w=400$~nm and $h_\text{mf}=0.1$~mT), and (b) the CPW signal line width $w$ ($x_0=0$ and $h_\text{mf}=0. 1$~mT). The gray area in (a) denotes region with $x_0 \in [0, 300~\mathrm{nm}]$, whereas in (b) are with $w>150$~nm.
}\label{fig:Fig4_v2} 
\end{figure}

The dependence of the SW intensity at $f_{0}, f_{2}, f_{3}$ on the position of the CPW ($w=400$ nm emitting a microwave magnetic field $h_\text{mf}=0.1$ mT) at $\mu_0 H_0=0.5$~T is shown in Fig.~\ref{fig:Fig4_v2}(a). Since $h_\text{mf}$ is below the threshold for $f_3$, the discussion below relates directly to $f_0$ and $f_2$, but similar dependencies were also found at higher values of $h_\text{mf}$ for $f_3$. As expected, the intensities of the SWs decrease as the CPW signal line is moved away from the centre of the cavity. The rate of decrease accelerates significantly at $x_0>425$ nm. This is due to the fact that, from this position, the emission occurs outside the cavity, and the excitation of the cavity mode is only via evanescent SW modes in the ADL at the frequency of 24.3~GHz, which is in the bandgap. The study shows that the $x_0\approx0$ is the most appropriate for excitation of SWs inside the cavity for CPW narrower than the cavity length. 

The influence of the width of the CPW signal line on the intensity of the excited SW modes is shown in Fig.~\ref{fig:Fig3}(b). We place the CPW centrally over the cavity and increase $w$ starting at $w=40$~nm, we keep $h_\text{mf}=0.1$~mT. The intensity of $f_0$ and $f_2$ mode increases with $w$, and as expected saturates when the signal line covers the entire cavity. We conclude that the optimal conditions for multi-frequency SW mode generation are when the cavity is completely covered by the signal line, regardless of its position relative to the cavity centre. However, if the signal line is narrower than the cavity, the intensity of the excited SW modes becomes sensitive to $x_0$ and $w$. Moreover, their values influence also the threshold field values for the generation of high-frequency harmonics. As shown in Fig.~\ref{fig:Fig3}(b) the threshold fields are 0.04 and 0.4~mT for CPW with $w=40$ nm (empty symbols), which are higher than for $w=400$~nm, (solid symbols), 0.01 and 0.1~T, respectively.

\section{Summary}

In this article, we describe a method to enhance the SW energy confined in a magnonic crystal cavity created in a thin Py-film based ADL to enter the nonlinear regime. We have shown that the microwave-pumped fundamental cavity mode, which has the frequency of the plain film FMR fitting into a complete bandgap, can easily reach the nonlinearity threshold leading to the generation of modes at multiple frequencies. Thus, we propose a way to generate high frequency SWs in the range of 28 to 79~GHz and 42 to 119~GHz with very short wavelengths of 62 to 31~nm and 40 to 22~nm for the 2$^\text{nd}$ and 3$^\text{rd}$ harmonics, respectively, by varying the external magnetic field from 1 to 0.3~T. The harmonics follow known dependencies on the amplitude of the pumped microwave field and the magnitude of the static magnetic field\cite{Gurevich1996}, allowing the effectiveness of multi-frequency mode generation to be significantly enhanced. The effects demonstrated in this paper can be achieved in standard ADLs using a simple CPW antenna, whose main limitation is the overlap of the signal line with the cavity surface. 
This provides an opportunity to investigate magnonics applications at high frequencies, which overlap with 6G microwave bands and have very short wavelengths, up to 100,000 times shorter than the wavelength of microwaves at these frequencies.

Further optimization of the magnonic crystal structure can even increase the demonstrated conversion efficiency of FMR oscillations to high-frequency oscillations. This can be achieved if, instead of using a cavity to confine only the SW at the input frequency (fundamental cavity mode), the structure also allows the mode to be confined at the output frequency(s) (high-frequency harmonics). Learning from photonics, not only can the power consumption be greatly reduced, but in principle 100\% conversion can be achieved \cite{Bravo-Abad2007,Rodriguez2007}.

\subsection*{Acknowledgements}
This work was supported in part by the National Science Center Poland project OPUS-LAP no 2020/39/I/ST3/02413. The simulations were partially performed in Poznań Supercomputing and Networking Center and at the National Institute of Technology Calicut, India.


\section*{Methods}
\subsection*{Micromagnetic simulations}

To achieve precise outcomes for the ADL's dispersion relations and to gain a better understanding of the complexity of the SW modes inside the crystal cavity, we utilized FEM and FDTD method to solve the Landau-Lifshitz and Maxwell equations. Within simulations, every magnetic moment in the predefined unit cells is given by normalized unit vectors $\textbf{m}=\textbf{M}/M_\text{S},$ where $\textbf{M}$ is the total magnetization, and $M_\text{S}$ is the saturation magnetization of the ferromagnetic material. Our approach then focuses on solving the Landau-Lifshitz-Gilbert (LLG) equation:
\begin{equation}
\frac{d\textbf{m}}{dt}=\gamma\frac{1}{1+\alpha^2}\left(\textbf{m}\times \textbf{B}_\text{eff}+\alpha\left(\textbf{m}\times\left(\textbf{m}\times \textbf{B}_\text{eff}\right)\right)\right),
\label{Eq:LLGE}
\end{equation}
 where $\gamma$ is the gyromagnetic ratio and $\alpha$ is the dimensionless damping coefficient. The effective magnetic flux density field, $\textbf{B}_\mathrm{eff}$ includes externally applied field, $\textbf{B}_\mathrm{0}\equiv \mu_0 \textbf{H}_0$, together with the magnetostatic demagnetization, $\textbf{B}_\mathrm{d}$, and the exchange field, $\textbf{B}_\mathrm{exch}$:
\begin{equation}
\textbf{B}_\text{eff}=\textbf{B}_\text{ext}+\textbf{B}_\text{d}+\textbf{B}_\text{exch}.
\label{Eq:b_eff}
\end{equation}

\subsubsection*{Finite element method}

In the FEM analysis implemented with Comsol Multiphysics, we addressed the eigenproblem derived from Eqs.~\ref{Eq:LLGE} and \ref{Eq:b_eff}, neglecting damping, i.e., $\alpha=0$, and defining the demagnetizing field amplitude, $\textbf{H}_\text{d}$ ($\equiv\textbf{B}_\mathrm{d}/\mu_{\text{0}}$), as the gradient of the magnetic scalar potential, $U_\text{m}$: $\textbf{H}_\text{d}=-\nabla U_\text{m}$, which inside a magnetic body fulfill the equation: $\nabla^2 U_\text{m}=\nabla\cdot\textbf{M}$. By considering full magnetization saturation along the direction $\hat{i}$ and utilizing a linear approximation, we could separate the magnetization vector into its static and dynamic (dependent on time $t$ and position $\textbf{r}$) components as $\textbf{m}(\textbf{r},t) = m_i \hat{i} + \delta \textbf{m}(\textbf{r},t)\;\forall\;(\delta \textbf{m}\perp\hat{i})$, neglecting all nonlinear terms in the dynamic magnetization $\delta\textbf{m}(\textbf{r},t)$. Additional information on this methodology can be found in Refs.~[\onlinecite{Mruczkiewicz2013StandingCrystals,Rychy2018SpinRegime}].

Using Bloch-Floquet boundary conditions (BC) on the unit cell boundaries:
\begin{equation}
 \textbf{m}_\text{dst}=\textbf{m}_\text{src}e^{-i\textbf{k}\cdot(\textbf{r}_\text{dst}-\textbf{r}_\text{src})},
\end{equation}
we model an infinite MC layer. Here, $\textbf{k}$ is the 2D Bloch wavenumber, $\textbf{m}$ is the normalized magnetic moment defined  both at the target (dst) and source (src), $\textbf{r}$ are the spatial coordinates of the boundaries where the BCs are applied, and $i$ is the imaginary unit. By parametrically sweeping the wavenumber, eigenfrequencies are determined at each interval. The resulting wavenumber versus frequency plots yields the dispersion curves for the periodic structure~\cite{Hakoda2018floquet, Collet2011floquet}.

For planes perpendicular to the plane of the MC layer, we have used Dirichlet BCs to suppress the scalar magnetic potential ($U_\text{m}$) at its out-of-plane boundaries. To ensure the physical accuracy and convergence of the simulation, it is crucial to position the conditions sufficiently far from the specimen. In our simulations, we set the cell's height to be 10 times the layer's unit cell width (1.5~\textmu m). The average quality of the tetrahedral discretization mesh, as determined by the volume-to-length parameter, is 0.91. This results in a mesh comprising 13630 prisms for the MCs unit cell, with one element per thickness and excluding the surrounding environment. The cavity mode simulations in Comsol Multiphysics (Fig.~\ref{fig:Fig6}) used the unit supercell approach, which consists of a cavity surrounded by 6 (parallel to the long side) and 8 (perpendicular) antidotes. This results in a symmetrical size of the supercell with a side equal to 1.35~\textmu m. In this case, the height of the environment is 20 times this width (27~\textmu m), and the amount of mesh elements of this supercell is 37330 with an average quality of 0.82 (and with one prism per thickness).

\subsubsection*{Finite difference time domain method}
The time domain simulations in the paper have been obtained with the open-source software MuMax3 employing the finite difference time domain method \cite{vansteenkiste2014design} with the RK45 solver (based on Dormand-Prince method) \cite{dormand1980family} to solve Eq.~(\ref{Eq:LLGE}). The micromagnetic cell size was taken as $5\times5\times5$~nm$^{3}$ with cell dimensions smaller than the exchange length $\left(l_{\text{ex}}=5.7\,\text{nm}\ \text{for Py}\right)$. We used the in-plane periodic boundary condition (PBC), i.e., the supercell of dimension 4.5~\textmu m $\times$ 3~\textmu m has been multiplied 10 and 50 times along the $x$ and $y$ directions, respectively.

To get the mode profiles, we calculate each magnetization component's pointwise FFT over time and then visualize the real part corresponding to a particular resonance frequency, see details in Ref.~[\onlinecite{gruszecki2023}].

The nonlinear response of the system excited by a single frequency microwave source was analyzed accordingly with the methodology used in Ref.~[\onlinecite{gruszecki2022}]. 

The mf current transmitted along the $y$ axis generates a magnetic field $\mathbf{h}_{\mathrm{mf}}$, which exerts a torque on the magnetization in Py. The dependence of $\mathbf{h}_{\mathrm{mf}}$ with peak amplitude $h_\text{mf}$ on $x$ coordinate can be approximated by the equation\cite{gruszecki2016microwave}:
\begin{equation}
\mathbf{h}_{\mathrm{mf}}(x)= h_{\mathrm{mf}}
\begin{bmatrix}
\frac{1}{2} + \frac{1}{2}\mathrm{cos}\left[\frac{2\pi}{2s} (|x|-\frac{w}{2})\right] \\ 
0 \\ 
\frac{1}{5} \mathrm{sign}(x)
\left\{
\frac{1}{2} + \frac{1}{2} \mathrm{cos}\left[\frac{2\pi}{s}(|x|-\frac{w}{2}-\frac{s}{2}\right] 
\right\}
\end{bmatrix}.
\label{eq:h_rf}
\end{equation}

\subsection*{DATA AVAILABILITY}
The data that support the findings of this study are available from the corresponding author upon reasonable request.

\bibliography{literature}
\end{document}